\documentclass{PoS}

\makeatletter
\renewcommand\@biblabel[1]{}
\makeatother

\def\ga{\mathrel{\raise0.35ex\hbox{$\scriptstyle >$}\kern-0.6em
\lower0.40ex\hbox{{$\scriptstyle \sim$}}}}
\def\la{\mathrel{\raise0.35ex\hbox{$\scriptstyle <$}\kern-0.6em
\lower0.40ex\hbox{{$\scriptstyle \sim$}}}}
\def\co{CO {\it J}=1-0 }
\def\cotwo{CO {\it J}=2-1 }

\def\cosix{CO {\it J}=6-5 }

\def\hcn{HCN {\it J}=1-0 }
\def\hcop{HCO$^+$ {\it J}=1-0 }

\def\cs{CS {\it J}=1-0 }
\def\hij{high-{\it J} }
\def\loj{low-{\it J} }

\title{Enabling the next generation of cm-wavelength studies of high-redshift molecular gas with the SKA}

\ShortTitle{Molecular gas studies with the SKA}

\author{
\speaker{Jeff Wagg}$^1$, Elisabete Da Cunha$^2$, Christopher L.
Carilli$^{3,4}$, Fabian Walter$^2$,  Manuel Aravena$^5$, Ian Heywood$^{6,7}$, Jacqueline
Hodge$^8$, Eric Murphy$^9$,  Dominik Riechers$^{10}$, Mark
Sargent$^{11}$, and Ran Wang$^{12}$
\\
$^1$SKA Organisation, Lower Withington, UK; 
$^2$MPIA, Heidelberg, Germany; 
$^3$NRAO, Socorro, USA;
$^4$Cavendish Astrophysics Group, Cambridge, UK;
$^5$U. Diego Portales, Santiago, Chile;
$^6$CASS, Sydney, Australia;
$^7$RATT, Dept. of Physics and Electronics, Rhodes University, Grahamstown, South Africa;
$^8$NRAO, Charlottesville, USA;
$^9$IPAC, Caltech, Pasadena, USA;
$^{10}$Cornell University, Ithaca, USA;
$^{11}$Astronomy Centre, Department of Physics and Astronomy, University of Sussex, Brighton, UK;
$^{12}$KIAA, Peking, Beijing, China
\\

E-mail: \email{j.wagg@skatelescope.org}
}

\abstract{
The Square Kilometre Array will be a revolutionary instrument for the
study of gas in the distant Universe. SKA1 will have sufficient sensitivity to detect and image
atomic 21~cm HI in individual
galaxies at significant cosmological distances, complementing ongoing ALMA imaging of redshifted \hij CO line emission
and far-infrared interstellar medium lines such as [CII]
157.7$\mu$m. At frequencies below $\sim$50~GHz, observations of redshifted emission from
\loj transitions of CO, HCN, HCO$^+$, HNC, H$_2$O and CS provide
insight into the kinematics and mass budget of the cold, dense star-forming
gas in galaxies. In advance of ALMA band 1 deployment (35 to 52~GHz),
the most sensitive facility for high-redshift studies of molecular gas
operating below 50~GHz is the Karl G. Jansky Very Large Array (VLA).   
Here, we present an overview of the role that the SKA could play in
molecular emission line studies during
SKA1 and SKA2, with an emphasis on studies of the dense gas
tracers directly probing regions of active star-formation.}

\FullConference{
Advancing Astrophysics with the Square Kilometre Array\\
June 8-13, 2014\\
Giardini Naxos, Italy}

\newcommand{\skipthis}[1]{}
\newcommand\nar{New Astronomy Reviews}
\newcommand\apj{ApJ}

\begin{document}

\section{Gas in galaxies: 2020 and beyond}

Studies of the dense star-forming gas in high-redshift galaxies have
been transformed in recent years owing to the increased
sensitivity and bandwidth of new submm through cm-wavelenth facilities
like the Karl G. Jansky Very Large Array (VLA), the Plateau de Bure
Interferometer (PdBI) and the Atacama Large Millimeter/submm Array
(ALMA). A few highlights include observations of molecular gas
through $^{12}$CO (hereafter CO) line emission in main sequence star-forming galaxies at $z
\sim 1.5 - 2.0$ (Daddi et al.\ 2008, 2010; Dannerbauer et al.\ 2009;
Tacconi et al.\ 2010, 2013;
Aravena et al.\ 2010, 2014), the detection and
imaging of CO and [CII] line emission in quasar host galaxies during
the epoch of reionization (Walter et al.\ 2003, 2004; Carilli et al.\
2007; Wang et al.\ 2013; 
Willott et al.\ 2013), and the first blank-field CO emission line surveys for gas-rich
galaxies over significant cosmological volumes (Walter et al.\ 2014;
Decarli et al.\ 2014). These new surveys greatly expand our
knowledge of the gas content of galaxies in the distant Universe, which had
been based primarily on detections of CO in far-infrared luminous AGN host
galaxies and submm/mm-selected star-forming galaxies (see Carilli \& Walter 2013 for a recent
review). The emerging picture is that, with the exception of atomic
HI, the cool gas content of main sequence star-forming galaxies is now accessible with current
facilities across
the peak in cosmic star-formation. Here, we summarize the 
molecular gas studies of high-redshift galaxies that could be enabled
by the SKA.

In 2020 and beyond, the SKA will operate in parallel with major new facilities covering much of the electromagnetic
spectrum, collaboratively conducting sensitive studies of gas and star-formation in
galaxies well into the epoch of reionization (EoR) at $z > 6$. These
facilities include optical telescopes like the European Extremely Large Telescope (E-ELT), the
Thirty Meter Telescope (TMT), and the Large Synoptic Survey Telescope
(LSST), and infrared facilities such as the \textit{James Webb Space Telescope}
(\textit{JWST}) and \textit{Euclid}. At long submm/mm-wavelengths, ALMA, NOEMA, the
Cerro Chajnantor Atacama Telescope (CCAT) and the Large Millimeter
Telescope (LMT) will shed light on Galactic and extragalactic regions of the 
Universe obscured by molecular gas and dust. The long wavelength
complement to these facilities will be the SKA.

One of the main science drivers for the SKA has always been the study
of 21~cm HI
line emission in normal galaxies out to redshifts $z \ga 1$. Predictions and science drivers for future SKA extragalactic HI surveys are
presented by other authors in this volume (e.g. Blyth et al.; de~Blok
et al.; Morganti et al.;
Obreschkow et al.; Santos et al.; Staveley-Smith et al.). Similarly, the SKA will
provide the unique ability to survey thermal and non-thermal (synchrotron) radio emission at low
radio frequencies covering 10s to 1000s of square degrees in order to probe the evolution of
obscured star-formation out to redshifts approaching the end of
reionization at $z \sim 6$ (see Prandoni et al., Jarvis et al., and Murphy
et al. in this volume). Over fields much smaller than one square degree, this obscured
star-formation will be probed to higher redshifts by ALMA,
which has already been successful in imaging far-infrared
continuum emission in star-forming galaxies during the epoch of
reionization (e.g. Wang et al.\ 2013). For galaxy evolution studies, the wide field imaging capabilities of the SKA will be
complemented at optical/infrared wavelengths by LSST and \textit{Euclid}. 
Finally, ALMA is sensitive to the \hij CO, HCN, HCO$^+$ and CS line
emission in the distant Universe at frequencies $\nu > 86$~GHz ($\nu >
35$~GHz once band 1 is deployed), leaving open the need for a
sensitive facility able to conduct surveys of the \loj transitions of
these molecules that provide an anchor for the spectral line energy
distribution. Without this anchor, there will be degeneracies in the
models fit to derive gas properties such as density and temperature. 
Measurements of only the \hij CO emission lines may lead
to a bias in molecular gas mass estimates as their intensities will
depend on the excitation conditions of the gas.

\section{Molecular CO line emission with the SKA}

\begin{figure}[h]

\centerline{\includegraphics[width=0.9\columnwidth]{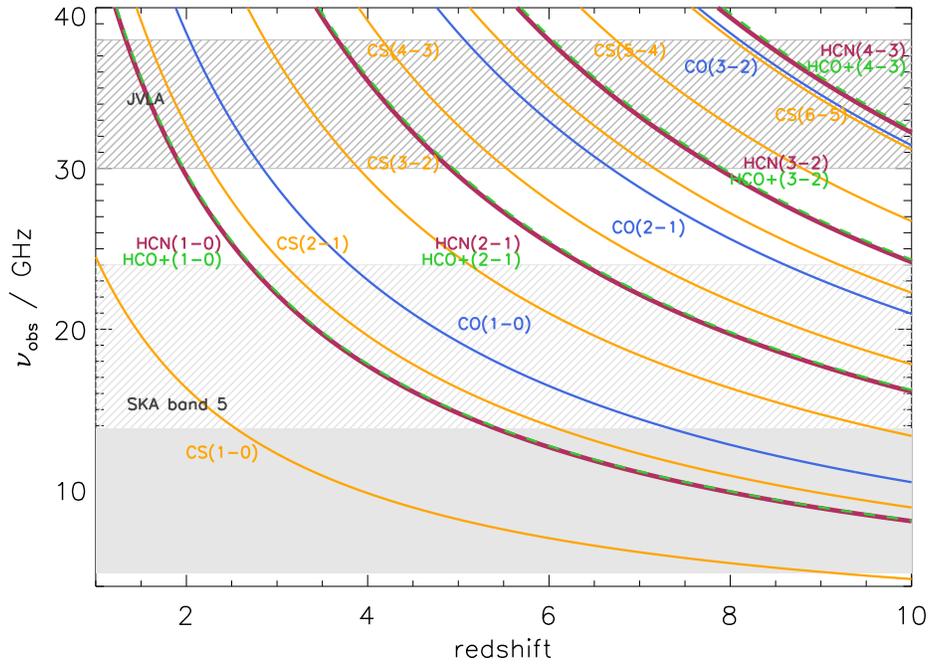}}

\caption{Redshifted line frequencies for multiple \loj transitions of
  CO, CS, HCN, and HCO$^+$. The \textit{grey} region shows the
  currently defined frequency range of band 5 on SKA1-MID, and the
  proposed extension to 24~GHz for SKA2 is shown as a \textit{hatched}
  region. At higher frequencies we show the frequency coverage provided by an
  8~GHz bandwidth VLA survey with the Ka-band receivers (26.5 to
  40~GHz). ALMA band 1 is expected to cover 35 to 52~GHz (Di~Francesco et
  al.\ 2013).}
\label{fig1}
\end{figure}

Among the outstanding challenges for galaxy formation studies is to
quantify the kinematic properties and to measure the gas content of
star-forming galaxies over the history of the Universe. Aided by the typical increase in line flux density with increasing rotational
transition in star-forming molecular gas,  interferometric observations of CO line emission provide a means of
estimating dynamical masses through the linewidths and source sizes in
galaxies  out to the very early Universe (e.g. Walter et al.\ 2004;
Riechers et al.\ 2013). At cm-wavelengths, one is sensitive to the \loj transitions of
this line, the luminosity of which has traditionally been used as
a means of estimating the total cold molecular ($= H_2$) mass that is
available to fuel star-formation by adopting a CO-to-H$_2$ conversion
factor, $\alpha_{CO}$ (e.g Downes \& Solomon 1998; Bolatto et al.\ 2013). As such, observations of high-redshift molecular CO line
emission have naturally become one of the main science drivers for current and future
cm-wavelength interferometers operating above $\sim$20~GHz. Figure~1
shows the redshifted frequencies of the \loj CO lines, along with
those of dense gas tracing molecular line species.

Here, we consider the likelihood that \co line emission will be either
a viable tracer of galaxy dynamics or a tool for estimating cold molecular
gas masses with the current plans for SKA1-MID and SKA2. The
possibility of CO intensity mapping with the SKA is discussed in
another chapter (Chang et al.). Five bands are currently
defined for SKA1-MID, although only three are expected to be
deployed in the first phase. Band 5 would extend up to 13.8~GHz, corresponding to \co
line emission redshifted to $z = 7.35$. The most luminous
metal-enriched quasar host galaxy at $z = 7.1$ is undetected in
\cosix line emission with the PdBI (Venemans et al.\ 2012), which should benefit from a factor
of $\sim$36$\times$ increase in flux density with respect to the \co
line emission owing to the $\nu^2$ dependance of flux density on
rest frequency. Molecular line emission from transitions where the
excitation temperature is 
similar to the CMB background temperature ($\rm T_{CMB}(z) = 2.7 (1 + z)$)
should not be detectable (e.g. Papadopoulos et al.\ 2000; Obreschkow
et al.\ 2009; Da Cunha et al.\ 2013b). In the case of
a $z = 7.5$ star-forming galaxy (similar to a submm galaxy or a far-infrared
luminous quasar host galaxy) with molecular gas at an average 
kinetic temperature of $\rm T_k = 40$~K (at $z = 0$), the \co line is predicted to exhibit
$\sim$40\% of the flux density it would have in the absence of the
CMB. In the case of more quiescent star-formation where the kinetic temperature of the
gas would be $\rm T_k = 18$~K , the integrated \co line flux
density of that galaxy at $z = 7.5$ would be down to 5\% of its $z =
0$ value. Such effects likely contributed to the (sensitive) Green
Bank Telescope (GBT) non-detections of \co
in two star-forming Ly$\alpha$ emitters (LAEs) at $z > 6.5$ by Wagg,
Kanekar \& Carilli (2009), including the gravitationally lensed LAE
HCM6A, which may have a star-formation rate
$\ga 100$~M$_{\odot}$~yr$^{-1}$ (Chary et al.\ 2005).     
During these early cosmic times, corresponding to the epoch of
reionization (EoR), measuring the kinematic properties (dynamical mass estimates)
of quasar host galaxies, or redshifts for the lower luminosity galaxy population, would
best be achieved through observations of the redshifted [CII] line
emission at mm-wavelengths (e.g. Maiolino et al.\ 2005; Walter et al.\
2009; Wagg et al.\ 2012a; Carilli et al.\ 2013; Wang et al.\ 2013;
Riechers et al.\ 2014).

Adding to the inherent difficulty in detecting the \loj CO lines
during the EoR, the average metallicity of galaxies is expected to be
lower, and recent observational and theoretical work suggests that
$\alpha_{CO}$ may be higher in low metallicity gas clouds
(e.g. Genzel et al.\ 2012; Leroy et al.\ 2009; Narayanan et al.\ 2012). Larger $\alpha_{CO}$
naturally leads to a lower CO line flux density per unit mass of
H$_2$. The $z  > 6$ population of star-forming Ly$\alpha$ emitters
 have typical star-formation rates of
5-10~$M_{\odot}$~yr$^{-1}$. Their metallicities cannot be measured with
current facilities (except possibly with ALMA, which is sensitive to
redshifted FIR lines like [CII] and [NII]), although these are likely
to have sub-solar metallicties given the timescales needed for metal
enrichment (the Universe is $\sim$710~Myr old at $z = 7.5$). 
Wagg et al.\ (2012b) obtained a sensitive GBT upper-limit on 
  \cotwo line emission in the Ly$\alpha$ blob (LAB) `Himiko' at $z =
  6.6$. It is likely that the low expected metallicities of typical $z > 6$
  star-forming galaxy populations like the LABs 
and LAEs will mean that SKA1-MID band 5 surveys of \co line emission will
only be sensitive to more massive and rare starburst galaxies (see
Murphy et al. in this edition). The only high-redshift galaxy
currently detected in molecular CO line emission at frequencies below 20~GHz is a
gravitationally lensed star-forming submm galaxy at $z = 6.34$ (Riechers
et al.\ 2013). The surface density of similarly luminous quasar host
galaxies detected in CO line emission at $z >  6$ is $\sim$1 per 500
square degrees.

During SKA2, the situation would be improved by the expected
order of magnitude increase in collecting area and extended frequency coverage up to 24~GHz (or
30 GHz, as proposed by Murphy et al.\  in this volume, who also
present the science case for  $\nu > 10$~GHz extragalactic continuum studies). Such studies
would be sensitive to \co line emission at  $z > 3.8$ ($z >
2.8$ for $\nu < 30$~GHz), covering the epoch when massive $z \sim 2$
galaxies should have formed their stars, likely in obscured bursts of
star-formation. For the case of cold, quiescent gas at a kinetic
temperature of 18~K in a $z > 4$ galaxy, the prediction is that
the measured flux density should be less than 25\% of what would be
measured in the absence of the CMB (Da~Cunha et al.\ 2013b). For warm
star-forming gas at $\rm T_k = 40$~K, one would measure $\sim$65\% of the
$z = 0$ intensity at a redshift of $z = 4$. We expect that SKA2-MID should
provide a powerful means of quantifying the cosmic evolution of the
molecular H$_2$ gas density over the epoch of massive galaxy
formation.

\section{Dense star-forming gas}

Dense star-forming gas tracers such as HCN and CS hold the promise for
future SKA studies of the high-redshift interstellar medium (ISM), as their
emission lines directly probe
sites of active star-formation with densities in excess of
10$^4$~cm$^{-3}$. Although they are typically an order of magnitude
fainter than the CO lines, the luminosity in the 88.6~GHz \hcn line 
correlates linearly with infrared luminosity in star-forming gas over
nearly eight orders of magnitude in luminosity
(Gao \& Solomon 2004; Wu et al.\ 2005). As such, emission from lines
like \hcn should be a good proxy for the total mass in dense molecular gas
directly involved in ongoing star-formation activity. 
The \hcn line traces gas at kinetic
temperatures of 40 to 50~K in luminous infrared star-forming galaxies, much
higher than the CMB temperature at redshifts below $z  \sim 6$. 
Current cm-wavelength facilites lack the sensivity to
detect \hcn line emission at $\nu \la 24$~GHz  in even the most
luminous, known star-forming galaxies and quasar host
galaxies. Previous attempts to detect \loj HCN line emission in $z >
2$ galaxies have only been successful for a handful of lensed objects
(Solomon et al.\ 2003; Vanden~Bout et al.\ 2005; Carilli et al.\ 2005;
Gao et al.\ 2007; Riechers et al.\ 2007), while \hij HCN line emission remains a promising tool
for ALMA studies (Wagg et al.\ 2005; Weiss et al.\ 2007; Danielson et al.\ 2011). The
HCO$^+$ line may also hold promise as a dense gas tracer in
high-redshift galaxies  (e.g. Riechers et al.\ 2006; Garcia-Burillo et
al.\ 2006), although the possible decrease in abundance of HCO$^+$ in regions of high electron
density has brought into question its effectiveness as a proxy for dense
molecular gas mass in starburst galaxies (e.g Papadopoulos et al.\ 2007).

In the case of CS, the \textit{J}=1-0 transition occurs at a rest
frequency of 49~GHz, close to the atmospheric O$_2$ line, which limits the
sensitivity for surveys of CS line emission in the local
Universe. CS is a more common tracer of dense star-forming
cores in our Galaxy (e.g. Evans 1999), however it has not been widely
studied in the high-redshift Universe given the sensitivity of current
cm-wavelength instruments.

If band 5 were deployed on SKA1-MID, it would be sensitive to \hcn
and \hcop line emission at $z > 5.4$ and CS~\textit{J}=1-0 line
emission at $z >  2.6$. An even greater advance will be made 
with the increased sensitivity and extended frequency coverage of
SKA2, which should be an excellent facility for the study of \loj transitions of dense gas
tracers in high-redshift galaxies. It would be the first interferometer
capable of resolving the emission from the \loj dense gas tracers at redshifts between 
$z \sim 1$ (roughly half the present-day age of the Universe), and to
the end of the epoch of reionization at $z \sim 6$. In Section~5, we
make predictions for the expected number of dense gas emitters that
might be detected in future SKA1 and SKA2 spectral line surveys.

\section{OH and H$_2$O megamaser emission}

Another potentially interesting probe of star-formation and the
molecular ISM of high-redshift galaxies is redshifted emission from OH
or H$_2$O megamaser emission. Although it may be that the luminosity in the OH
megamaser line is correlated with star-formation activity, this
emission has not been detected in far-infrared luminous galaxies at
high-redshift (e.g. Ivison 2006). Similarly, early searches for H$_2$O
megamasers at high-redshift (e.g. Wilner et al.\ 1999; Ivison 2006;
Edmonds et al.\ 2009; Wagg \& Momjian 2009), resulted in only a pair of detections at $z =
0.66$ (Barvainis \& Antonucci 2005) and $z = 2.64$ (Impellizzeri et
al.\ 2008). It is therefore difficult to predict the expected number
density of these line emitters with the increased sensitivity of the
SKA. Murphy et al. (this edition) discuss the use of H$_2$O megamaser
line emission for studying AGN at high-redshift.

\section{Predictions for SKA spectral line surveys}

\begin{figure}[h]

\centerline{\includegraphics[width=0.9\columnwidth]{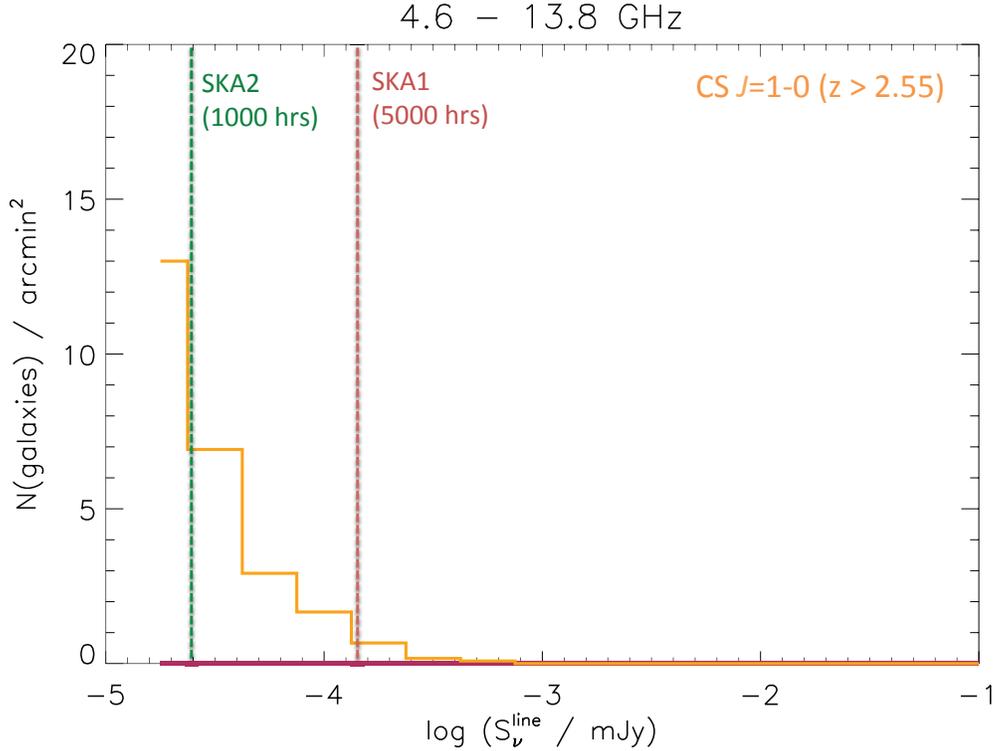}}

\caption{Predicted number density of \cs line emitters (orange curve) as a function
  of peak flux density, assuming a 300~km~s$^{-1}$ linewidth. In a
  deep SKA1 survey (5000~hours integration time), one would expect to detect
  approximately one source per square arcminute over the field of view at 13.8~GHz. That number decreases to zero for SKA1
  with half the sensitivity of the baseline design, but increases to
   10 to 13 per square arcminute for an SKA2 survey (1000~hours integration time). }
\label{fig2}
\end{figure}

For the purpose of predicting the expected detectability of molecular
emission lines in future SKA observations, we first assume a
$\Lambda$-dominated cosmological model with $\Omega_{\Lambda} = 0.7$,
$\Omega_{M} = 0.3$, and $H_0 = 70$~km~s$^{-1}$~Mpc$^{-1}$. The standard
definition of line luminosity is given by Solomon, Downes \& Radford
(1992), and we assume a linewidth, $FWHM = 300$~km~s$^{-1}$
throughout. For both SKA1 and SKA2, we consider two scenarios, one
where targeted observations of known ultraluminous infrared galaxies (ULIRGs) are made ($L_{IR} =
10^{12}$~L$_{\odot}$, $SFR \sim 100$~M$_{\odot}$~yr$^{-1}$), and a
second where pointed deep field observations are conducted, sensitive to less
luminous galaxies. In the latter case, we adopt the models of da~Cunha
et al.\  (2013a), who make empirical predictions for the observed
molecular line properties of star-forming galaxies in the observed
\textit{Hubble} Ultra Deep Field (UDF). In the original models, the
expected CO line luminosity is predicted from the infrared luminosity,
and here we extend this prediction to HCN, HCO$^+$, and CS which we convert
from the \co line luminosity by adopting the relationship to dense gas
tracer line luminosity measured by Gao \& Solomon (2004). 

For the purposes of SKA1 band 5 (4.6 to 13.8 GHz) predictions, we do
not consider \co line emission at $z > 7.4$, due to the effects
discussed in the previous section. ALMA observations of
higher-$\textit{J}$ CO lines would be sensitive to the same
star-forming molecular gas, while also benefitting from an 
increase in flux density that depends on $\nu^2$. 
For targeted observations of 
\hcn or \cs line emission in ULIRGs at $z \sim 5.4$ (HCN) or $z \sim  2.6$
(CS), the expected peak flux density in both cases would be $\sim$1.5~$\mu$Jy. A
5-$\sigma$ detection of such a line would take $\sim$1000~hours of
integration time with the full SKA1, or $\sim$4000~hours if SKA1
had 50\% less sensitivity. Now considering predictions for blind
surveys, Figure~2 shows the expected number density
of \cs line emitters as a function of flux density, where we also
indicate the expected sensitity after 5000~hours with the full
SKA1. In that case, one might expect to detect $\sim 1$ source per
square arcminute over the field of view at the highest
frequency end of the band. 

\begin{figure}[h]

\centerline{\includegraphics[width=0.9\columnwidth]{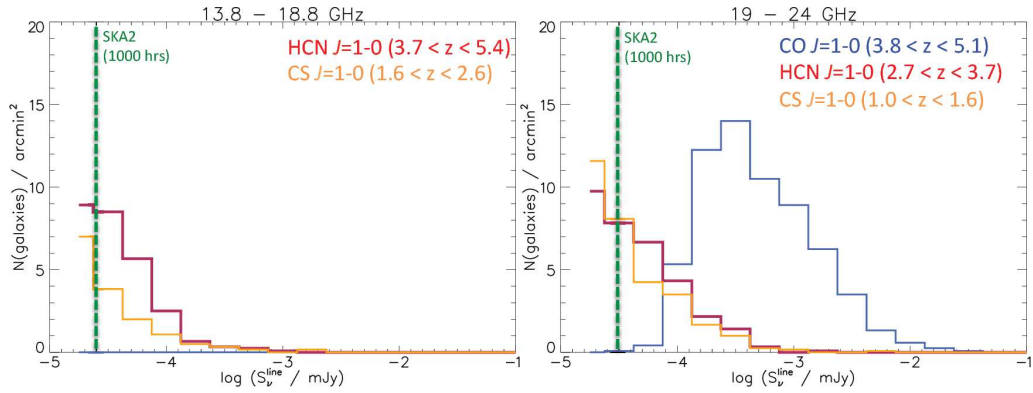}}

\caption{Predictions for the number density density of molecular line
  emitters as a function of peak line flux density (assuming a
  300~km~s$^{-1}$ linewidth). The \textit{left} panel shows the
  prediction for a 5~GHz wide tuning between 13.8 and 18.8~GHz, while
  the \textit{right} panel shows the same for a 5~GHz wide tuning
  extending up to 24~GHz. \cs line emitters are indicated by the
  orange curves, while the red curves show \hcn (or \hcop ) and the
  blue curve in the right-hand panel indicates the expected number density of
  \co line emitters. Note that the decline in the number density of
  low luminosity CO line emitters is artificial, as it is based
  on the observed number density of low luminosity optically selected galaxies in the
  UDF.}
\label{fig3}
\end{figure}

SKA2 should enable a revolution in high-redshift molecular gas
studies, not only because of the extended frequency coverage (up to 24~GHz), but also due to the
increase from 190 to 2500 in the number of band 5 MID antennas. In
assessing the potential for SKA2 studies,  we only assume blank-field searches for molecular line emitting
galaxies. We note that the expected rms in a 300~km~s$^{-1}$ channel after 1000~hours of
on-source integration time is expected to be $\sim$24~nJy at
frequencies above $\sim$10 GHz. Figure~3 shows the predictions for two tunings, each
covering 5~GHz of bandwidth over the upper-end of SKA2-MID band 5. We
only consider \co line emitters detected in the upper tuning ($z < 5$), when the
emission from cold molecular gas is not expected to be dramatically
effected by the CMB effects discussed previously. We note that at
these redshifts metallicity evolution may have an impact on the detectability of CO line emission
in galaxies with low metallicities. A deep integration covering the 19
to 24~GHz frequency range is expected to detect more than 50 CO line emitters per
square arcminute. Such a survey would provide the first strong
constraint on the evolution of the \co line luminosity function at $z > 4$.

\section{Summary}

The SKA has the potential to be a powerful facility for the detection
and imaging of \loj transitions of molecular line emission in
high-redshift galaxies. During 
SKA1, it could be possible to detect faint \cs line
emission associated with $z > 2.6$ luminous infrared galaxies
($L_{FIR} > 10^{12}$~L$_{\odot}$). At $z > 5.4$, \hcn could be observed
in order to quantify the total dense molecular gas mass in galaxies. Our models predict that with the
increased sensitivity and frequency coverage of SKA2 extending up to
$\sim$24~GHz, we expect to detect significant numbers of $z> 3.8$ CO line
emitters in blank-field surveys. Such galaxies may be significant
contributors to the total star-formation rate density of the Universe
at these early cosmic epochs.

\end{document}